\documentclass[]{JHEP3}

\usepackage{multicol}
\usepackage{epsf,amsmath,amssymb}

\newcommand{\code}{\tt}
\newcommand{\abbrev}{\small}
\newcommand{\ep}{\epsilon}

\newcommand{\api}{\frac{\alpha_s}{\pi}}

\newcommand{\eqn}[1]{Eq.\,(\ref{#1})}
\newcommand{\fig}[1]{Fig.\,\ref{#1}}

\newcommand{\reference}[1]{Ref.\,\cite{#1}}

\newcommand{\order}[1]{{\cal O}(#1)}

\newcommand{\lht}{l_{Ht}}
\newcommand{\Li}{{\rm Li}}

\newcommand{\lo}{{\abbrev LO}}
\newcommand{\nlo}{{\abbrev NLO}}
\newcommand{\nnlo}{{\abbrev NNLO}}
\newcommand{\gfermi}{G_{\rm F}}

\newcommand{\mhiggs}{M_H}
\newcommand{\pdf}{{\abbrev PDF}}
\newcommand{\mahiggs}{M_A}
\newcommand{\mssm}{{\abbrev MSSM}}
\newcommand{\lhc}{{\abbrev LHC}}
\newcommand{\coeff}{\tilde C}
\newcommand{\opo}{\tilde {\cal O}}
\newcommand{\bare}{{\rm B}}
\newcommand{\feynsl}[1]{
  \setbox0=\hbox{/} \setbox1=\hbox{$#1$}
  \dimen0=\wd0 \advance\dimen0 by -\wd1 \divide\dimen0 by 2
  \ifdim\wd0>\wd1 \lower.15ex
          \copy0\kern-\wd0\kern\dimen0\copy1\kern\dimen0
  \else \kern-\dimen0\lower.15ex
          \copy0\kern-\dimen0\kern-\wd1\copy1\fi}

\title{Production of a pseudo-scalar Higgs boson at hadron colliders at
  next-to-next-to-leading order}

\author{Robert V. Harlander\\
    {\it TH Division, CERN, CH-1211 Geneva, Switzerland}\\
    E-mail: \email{robert.harlander@cern.ch}}
\author{William B. Kilgore\\
    {\it Physics Department, Brookhaven National Laboratory,
      Upton, NY 11973, U.S.A.}\\
    E-mail: \email{kilgore@bnl.gov}}

\preprint{BNL-HET-02/17, CERN-TH/2002-186, hep-ph/0208096}

\abstract{ The production cross section for pseudo-scalar Higgs bosons
  at hadron colliders is computed at next-to-next-to-leading order
  (\nnlo) in {\abbrev QCD}.  The pseudo-scalar Higgs is assumed to couple
  only to top quarks. The \nnlo{} effects are evaluated using an
  effective lagrangian where the top quarks are integrated out. The
  \nnlo{} corrections are similar in size to those found for scalar
  Higgs boson production.}


\begin{document} 

\section{Introduction}
In its minimal version, the Standard Model of particle interactions
requires a complex scalar weak-isospin doublet to spontaneously break
the electro-weak gauge symmetry. Three of its four original degrees of
freedom transform into the longitudinal modes of the $W^\pm$ and $Z$
gauge bosons while the fourth manifests itself as a neutral scalar
field, the Higgs boson. Yukawa-type couplings between the Higgs field
and the fermions generate gauge invariant masses for the fermions.

There is no strong theoretical reason to believe that the minimal
version of the Standard Model is indeed realized in nature.  Extended
models give rise to several Higgs particles which can be
electrically charged or neutral and which can be even or odd
under {\abbrev CP} inversion.

The most appealing extension of the Standard Model, at the moment, is
the Minimal Supersymmetric Standard Model (\mssm{}), which basically
doubles the field content of the Standard Model and requires two Higgs
doublets for the generation of the particle masses.  This results in
five physical Higgs bosons: three neutral ($h,H,A$) and two charged
($H^\pm$).  $h$ and $H$ are {\abbrev CP}-even while $A$ is {\abbrev
CP}-odd (for a review see \reference{hunter}).

To date, no Higgs boson has been observed in nature, in spite of great
efforts that have been made at particle accelerators, especially at the
Large Electron Positron collider ({\abbrev LEP}) at {\abbrev CERN}. The
null results lead to lower limits on the mass of possible Higgs bosons.
For a scalar Higgs boson in the framework of the minimal Standard Model,
this limit is $\mhiggs\geq 114.4$\,GeV at $95\%$\,CL. Assuming the
\mssm{}, the limit for a neutral scalar Higgs goes down to about
91\,GeV. The $95\%$\,CL limit on a {\abbrev CP}-odd Higgs boson, the
subject of this paper, is around 92\,GeV\,\cite{lephiggs}.

The Large Hadron Collider (\lhc{}) at {\abbrev CERN} is scheduled to
commence taking data in the year 2007.  It will be a proton-proton
collider with a center of mass energy of $\sqrt{s} = 14$\,TeV and has
been designed specifically to search for Higgs bosons. For all
relevant values of the Higgs boson mass and most kinds of Higgs bosons
in various models, the gluon-gluon production mode will be of the greatest
importance both for discovery and for measuring the Higgs boson mass.
For {\abbrev CP}-even Higgs bosons, the theoretical prediction is now
fairly well under control, having been calculated to
next-to-next-to-leading order (\nnlo) in the strong coupling constant
\cite{HKggh,AnaMel}.

In this paper, we use the techniques of
\reference{virtual,HKggh,kichep02} to evaluate the production rate for
a pseudo-scalar Higgs boson to the same accuracy, i.e., to
next-to-next-to-leading order in {\abbrev QCD}.  We work in the
heavy-top limit, using an effective lagrangian for the interaction of
the pseudo-scalar Higgs boson with the gluons.  As in the case of the scalar
Higgs boson, this does not restrict the validity of the results to the
Higgs-mass region well below $2M_t$ if one factors in the
full top mass dependence at leading order \cite{nlo}.

In the effective lagrangian approach, the massive top-quark loop that
mediates the coupling between the Higgs and the constituents of the
initial state hadrons, reduces to effective vertices with known
coefficient functions. What remains to be computed at \nnlo{} are
$2\to 1$ processes up to two loops, $2\to 2$ up to one loop, and $2\to
3$ at tree level, where all internal particles are massless, and all
external particles are taken on-shell ($p_i^2=0$ for quarks and
gluons, $q^2=\mahiggs^2$ for the Higgs).

We note that we do not consider contributions from virtual particles
in extended theories in this paper.  We also do not consider the
effect of $b$ quark loops at this order, which can be important for
example in the \mssm{}, when the coupling is enhanced by a large value
of $\tan\beta$.  For light quarks, like the $b$, the effective
lagrangian cannot be formulated and one must perform a true three-loop
calculation at this order.  That calculation is beyond the current
state of the art.

\nlo{} corrections to this process were evaluated in
\reference{nlo,nloeff} and found to be very similar in size to the
\nlo{} effects for scalar Higgs production. We find that this is also
true at \nnlo{}: The K-factors at \nnlo{} for the scalar and the
pseudo-scalar are comparable.  This means in turn that the
production rate for pseudo-scalar Higgs bosons is fairly well under
control. Though still sizable, the \nnlo{} term in the perturbative
series is significantly smaller than the \nlo{} term and still higher order
effects are presumably negligible.  Accordingly, the unphysical
dependence of the cross section on the renormalization and
factorization scales is reasonably small, allowing for a prediction of
the total rate with errors of the order of, or below, the expected
experimental precision.

The paper is organized as follows: In the second section, we describe
our theoretical framework, including the effective lagrangian, its Wilson
coefficients and the renormalization of its operators, and our
prescription for handling Levi-Civita tensors and $\gamma_5$.  In the
third section, we briefly discuss the matrix element calculations and
the calculation of the partonic cross sections.  In section 4, we
present our results for the partonic cross sections.  The result for
pseudo-scalar production is very similar to that for scalar production
and the expression for the difference between the two is quite
compact.  The full expression is presented in the appendix.  In
section 5 we compute the hadronic cross section by folding in the
parton distributions and finally we present our conclusions.

\section{Theoretical setup}
In principle, the whole theoretical background is laid out very
clearly in \reference{CKSB}. Nevertheless, let us review the necessary
ingredients for our calculation.

We assume that the pseudo-scalar Higgs boson, $A$, couples only to top
quarks, $t$. The interaction vertex is given by
\begin{equation}
\begin{split}
  {\cal L}_{At\bar t} = -ig_t\frac{A}{v}\,M_t\,\bar t\gamma_5 t \,,
  \label{eq::ttA}
\end{split}
\end{equation}
where $g_t$ is a coupling constant that depends on the specific theory
under consideration. In the \mssm{}, for example, one has $g_t =
\cot\beta$, where $\tan\beta$ is the standard ratio of the vacuum expectation
values of the two Higgs fields.

The coupling of $A$ to two gluons is then mediated by a top quark loop.
In the heavy-top limit, this interaction can be described by an
effective lagrangian~\cite{CKSB}:
\begin{equation}
\begin{split}
  {\cal L}_{Agg} &= -g_t\,\frac{A}{v}\left[\coeff_1^\bare\, \opo_1^\bare
    + \coeff^\bare_{2}\,
    \opo^\bare_{2}\right]\,,\\
  \opo_1^\bare &= G^{a}_{\mu\nu}\tilde
  G^{a,\mu\nu}\,,\qquad \opo^\bare_{2} = \partial_\mu\left(
  \sum_{q}\bar q \gamma^\mu\,\gamma_5 q\right)\,.
  \label{eq::leff}
\end{split}
\end{equation}
$G_{\mu\nu}^a$ is the gluon field strength tensor, and  $\tilde
G^{a}_{\mu\nu}$ is its dual:
\begin{equation}
\begin{split}
  \tilde G^{a}_{\mu\nu} = \epsilon_{\mu\nu\alpha\beta}\,G^{a,\alpha\beta}\,.
  \label{eq::gdual}
\end{split}
\end{equation}
All quantities in these equations are to be understood as bare
quantities in the effective theory of five massless flavors. Thus the
sum over quarks ($\sum_q$) in \eqn{eq::leff} does not include the top
quark.  $\coeff_1^\bare$ and $\coeff^\bare_{2}$ are coefficient
functions that can be evaluated perturbatively. One may define
renormalized operators and coefficient functions as follows:
\begin{equation}
\begin{split}
\opo_i = \sum_{j=1}^2 Z_{ij} \opo_j^\bare\,,\qquad
\coeff_i = \sum_{j=1}^2 (Z^{-1})_{ji} \coeff_j^\bare\,,\qquad i = 1,2\,.
\label{eq::opren}
\end{split}
\end{equation}
The renormalization matrix $Z$ is given by~\cite{Larin}
\begin{equation}
\begin{split}
Z_{11} &= Z_{\alpha_s}\,,\qquad
Z_{12} = \frac{4}{\ep}\,\api +  \order{\alpha_s^2}\,, \\
Z_{21} 
&= 0\,,\qquad
\quad Z_{22} = Z_{\rm MS}^s\,Z_5^s\,,
\end{split}
\end{equation}
with
\begin{equation}
\begin{split}
Z_{\alpha_s} &= 1 - \api\,\frac{\beta_0}{\ep} 
+ \left(\api\right)^2\,\left(\frac{\beta_0^2}{\ep^2} 
  - \frac{\beta_1}{2\ep}\right) + \order{\alpha_s^3}\,,\\
\beta_0 &= \frac{1}{4}\left(11 - \frac{2}{3} n_f\right)\,,\qquad
\beta_1 = \frac{1}{16}\left(102 - \frac{38}{3} n_f\right)\,,\\
Z_{\rm MS}^s &= 1 + \left(\api\right)^2\,\frac{1}{\ep}\,\left[
   \frac{11}{6} + \frac{5}{36}\,n_f\right] + \order{\alpha_s^3}\,,\\
Z_5^s &= 1 - \api\,\frac{4}{3} + \order{\alpha_s^2}\,,
\label{eq::renconsts}
\end{split}
\end{equation}
where $n_f$ is the number of light (i.e.~massless in our approach) quark
flavors. In our numerical analysis below, we will always assume $n_f=5$.
$Z_{\alpha_s}$ and $Z_{\rm MS}^s$ are the renormalization
constants of the strong coupling and the singlet axial current,
respectively. $Z_5^s$ is a finite renormalization constant that will be
discussed below.

The coefficient functions are process independent and have been
evaluated in the context of pseudo-scalar Higgs decay to {\abbrev
  NNLO}~\cite{CKSB}:
\begin{equation}
\begin{split}
  \coeff_1(\alpha_s) &\equiv -\api\,\frac{1}{16} + \order{\alpha_s^4}\,,\\
  \coeff_{2}(\alpha_s) &= \left(\api\right)^2\,\left(
    \frac{1}{8} - \frac{1}{4}\,\ln\frac{\mu^2}{M_t^2}\right) + 
  \order{\alpha_s^3}\,,
\end{split}
\end{equation}
where $M_t$ is the pole mass of the top quark.

The matrix elements to be evaluated have the form
\begin{equation}
\begin{split}
\langle \coeff_1\opo_1 &+ \coeff_2\opo_2\rangle =
\coeff_1Z_{11}\,\langle \opo_1^\bare\rangle 
+ (\coeff_1Z_{12} + \coeff_2Z_{22})\,\langle \opo_2^\bare\rangle\,,
\label{eq::matel}
\end{split}
\end{equation}
with $\langle\opo_n^\bare\rangle \equiv \langle ab|
\opo_n^\bare|XH\rangle$, where $a$ and $b$ label the two partons in
the initial state, and $X$ denotes an arbitrary number of partons in
the final state.  We need to evaluate the square of this matrix
element up to $\order{\alpha_s^4}$.

\FIGURE{
    \leavevmode
    \begin{tabular}{cccc}
      \epsfxsize=8em
      \epsffile[180 550 340 680]{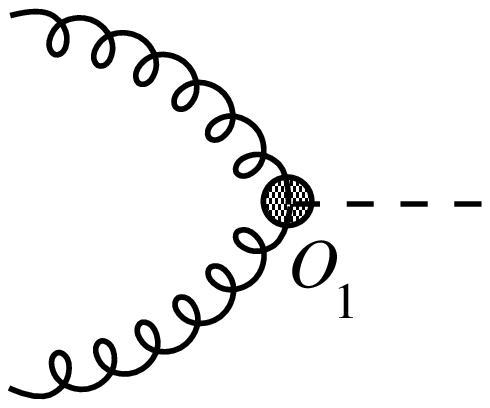}
      &
      \epsfxsize=8em
      \raisebox{1em}{
      \epsffile[180 580 350 670]{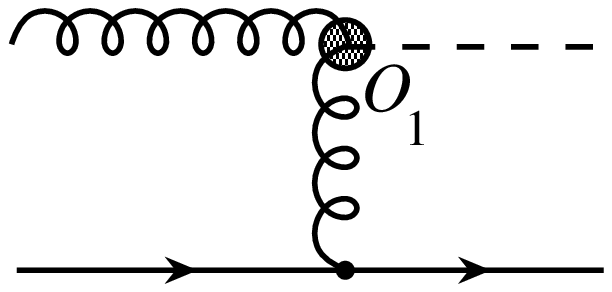}}
      &
      \epsfxsize=8em
      \epsffile[180 550 340 680]{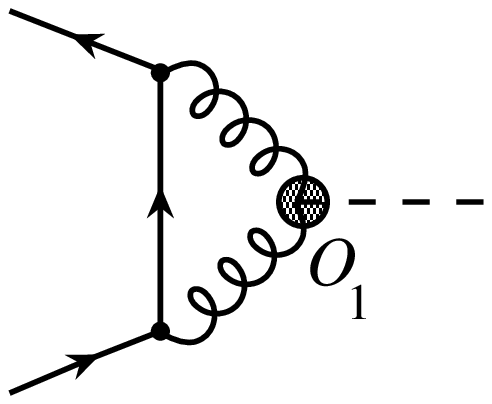}
      &
      \epsfxsize=10em
      \raisebox{.5em}{\epsffile[165 570 350 675]{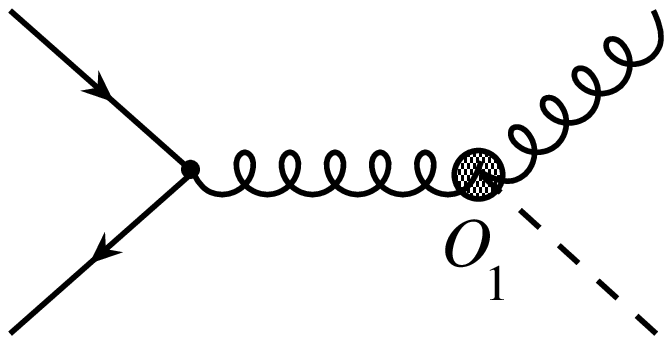}}\\
      $(a)$ & $(b)$ & $(c)$ & $(d)$
    \end{tabular}
      \caption[]{\label{fig::o1}\sloppy
        Sample diagrams for $\langle\opo_1^\bare\rangle$. Higher orders
        are obtained by dressing these diagrams with additional quarks
        and gluons, both virtual and real.
        }
}

\FIGURE{
    \begin{tabular}{cccc}
      \epsfxsize=8em
      \epsffile[180 550 340 680]{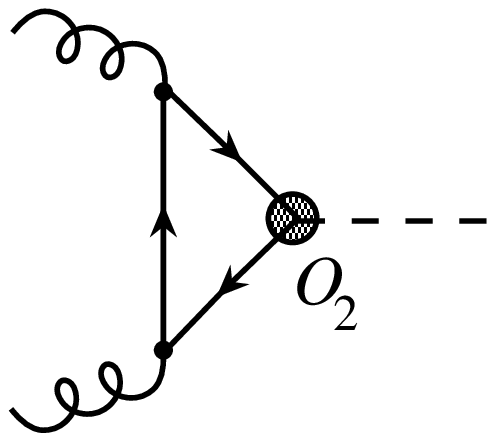}
      &
      \epsfxsize=8em
      \raisebox{1em}{
      \epsffile[180 580 350 670]{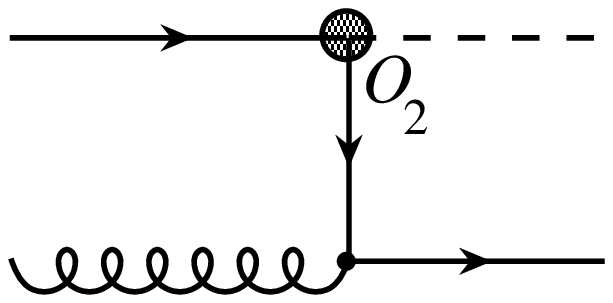}}
      &
      \epsfxsize=8em
      \epsffile[180 550 340 680]{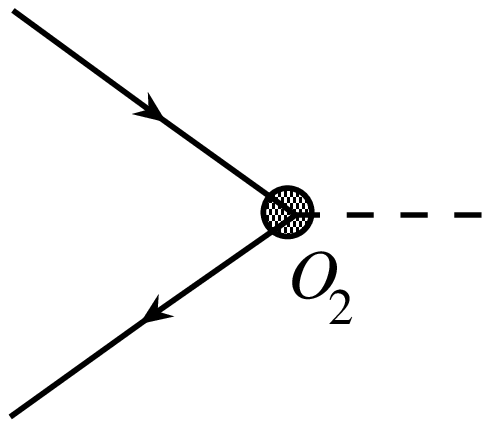}
      &
      \epsfxsize=8em
      \epsffile[180 550 340 680]{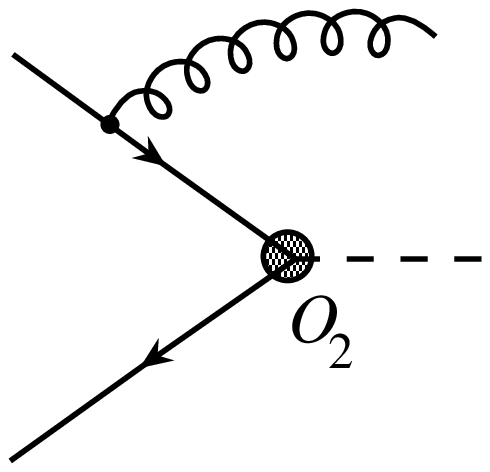}\\
      $(a)$ & $(b)$ & $(c)$ & $(d)$
    \end{tabular}
    \caption[]{\label{fig::o2}\sloppy
      Sample diagrams for $\langle\opo_2^\bare\rangle$ as required up to
      \nnlo{}.  }
}

$\opo_1^\bare$ generates vertices which couple two or three gluons to
the pseudo-scalar Higgs (the $ggggA$-vertex vanishes due to the
Jacobi-identity of the structure functions of SU(3)).  Thus, the
diagrams related to $\langle\opo_1^\bare\rangle$ are the same as in
the scalar Higgs case (cf.\,\reference{soft}). A
sample of typical diagrams is shown in \fig{fig::o1}. Higher order
contributions are obtained by dressing these diagrams with additional
gluons and quarks, either as virtual particles, or as real particles
in the final state.  The actual results for these diagrams are
different from the ones in the scalar case, of course, due to the
different Feynman rules corresponding to $G_{\mu\nu}G^{\mu\nu}$ and
$G_{\mu\nu}\tilde G^{\mu\nu}$.

As for $\langle\opo_2^\bare\rangle$, since its prefactor in
\eqn{eq::matel} is of order $\alpha_s^2$, it is only required to one
order less than $\langle\opo_1^\bare\rangle$.  Diagrams contributing to
$\langle\opo_2^\bare\rangle$ for the different subprocesses are shown in
\fig{fig::o2}. They appear in interference terms with the corresponding
diagrams of $\langle\opo_1^\bare\rangle$ (see \fig{fig::o1}).  At
\nnlo{}, the square of diagram $(c)$ is the only potential contribution
to the process arising solely through $\opo_2^\bare$. It vanishes,
however, in the limit that the light quarks are massless.

When evaluating the diagrams, the presence of the manifestly
4-dimensional Levi-Civita tensor in $\opo_1^\bare$ as well as the
$\gamma_5$ in $\opo_2^\bare$ requires special care.
We strictly follow the strategy outlined in \reference{Larin}.
This means that first we replace
\begin{equation}
\begin{split}
\feynsl{q}\gamma_5 &\to
\frac{i}{3!}q^{\alpha}\epsilon_{\alpha\beta\gamma\delta}\,
\gamma^\beta\gamma^\gamma\gamma^\delta\,.
\label{eq::g5def}
\end{split}
\end{equation}
Since we need to evaluate squared amplitudes, there will always be
exactly {\it two} Levi-Civita tensors in our expressions. This allows
us to express them in terms of metric tensors, which can be
interpreted as $d$-dimensional objects:
\begin{equation}
\begin{split}
\epsilon_{\alpha\beta\gamma\delta}
\epsilon^{\bar\alpha\bar\beta\bar\gamma\bar\delta}
= - g_{\alpha}^{[\bar\alpha}
g_{\beta}^{\bar\beta}
g_{\gamma}^{\bar\gamma}
g_{\delta}^{\bar\delta]}
= - g_{\alpha}^{\bar\alpha}g_{\beta}^{\bar\beta}
g_{\gamma}^{\bar\gamma}g_{\delta}^{\bar\delta} + 
g_{\alpha}^{\bar\beta}g_{\beta}^{\bar\alpha}
g_{\gamma}^{\bar\gamma}g_{\delta}^{\bar\delta} \mp \dots\,,
\end{split}
\end{equation}
where, as indicated, the brackets around the upper indices indicate
that their positions are to be anti-symmetrized while the lower
indices remain fixed.  All further calculations (i.e., integrations,
contraction of indices and spinor algebra) can be performed in $d$
dimensions.  This procedure is slightly different from that of
\reference{CKSB}, but we have checked that it gives the same result
for the pseudo-scalar Higgs decay up to
$\order{\alpha_s^4}$.\footnote{We thank M.~Steinhauser for providing
us with unpublished intermediate results concerning \reference{CKSB}.}
Furthermore, we have computed the two-loop virtual corrections for our
process in both approaches and obtained identical results.

The finite renormalization constant $Z^s_5$, introduced in
\eqn{eq::renconsts}, is determined by requiring that the one-loop
character of the operator relation of the axial anomaly is preserved
also at higher orders:
\begin{equation}
\begin{split}
\opo_2 = \api\frac{n_f}{8}\,\opo_1\,.
\label{eq::o1o2rel}
\end{split}
\end{equation}
Note that $\opo_1$ and $\opo_2$ denote {\it renormalized} operators
here, as defined in \eqn{eq::opren}.

\section{Methods of evaluation}

\subsection{Virtual corrections}
Potential sub-processes are $gg\to A$ and $q\bar q\to A$. It turns out,
however, that the latter does not contribute at \nnlo{}. The diagrams needed
are two-loop vertices with one massive and two massless external legs.

For their evaluation, we use the technique of \reference{baismi} that
has already been applied successfully to the evaluation of the virtual
correction to scalar Higgs production~\cite{virtual}. This means that
the two-loop vertex diagrams are first mapped onto three-loop
propagator diagrams by interpreting the massless external lines
(gluons or quarks) as part of an additional loop. The resulting
integrals, including tensor structures, can be treated by means of the
integration-by-parts algorithm of \reference{IP}. In particular, we
can use the {\code FORM}~\cite{form} program {\code
MINCER}~\cite{mincer} as the basis of an algebraic program that reduces
all integrals encountered to a set of master
integrals~\cite{virtual}. The analytic expressions for the latter have
been known for a long time~\cite{master}.  Note that a generalized
version of the method of \reference{baismi} has been constructed in
\reference{AnaMel}.

\subsection{Single real emission}
The radiation of one additional parton has to be evaluated up to
one-loop level. The contributing processes are $gg\to Ag$, $gq\to Aq$,
$g\bar{q}\to A\bar{q}$, and $q\bar{q}\to Ag$.  The one-loop matrix
elements have been evaluated to all orders in the dimensional
regularization parameter $\ep = (4-d)/2$.  After interfering with the
tree-level amplitude, the squared matrix element can be integrated
over single-emission phase space to obtain the contribution to the
partonic cross section in closed analytic form.  The interference of
bare operators $\opo_1^\bare$ and $\opo_2^\bare$ is of order $\ep$.
Nonetheless, operator $\opo_2^\bare$ contributes to the single real
emission cross section through operator mixing since $Z_{12}$ is of
order $1/\ep$.  The Feynman rules, loop integrals and phase space
integration have all been implemented in {\code FORM} programs.

\subsection{Double real emission}
We need the tree-level expressions for the processes $gg\to Agg$,
$gg\to Aq\bar q$, $gq\to Agq$, $q\bar q\to Agg$, $q\bar q\to Aq\bar
q$, and $qq\to Aqq$ (and the corresponding charged conjugated
processes).  The squared matrix elements can be evaluated
straightforwardly in $d=4-2\ep$ space-time dimensions with the help of
{\code FORM}.  The result is a rather large expression of several
thousand terms which must then be integrated over phase space.

The phase space for double real emission is quite complicated and we
perform the integration in the method of \reference{HKggh}.
That is, the matrix element and the phase space are expanded in terms
of $(1-x)$, where $x=\mahiggs^2/\hat s$. The result is a power series
expansion in $(1-x)$ and $\ln(1-x)$ (at \nnlo, the highest power of
the logarithm is $\ln^3(1-x)$) for the double real emission
contribution to the hadronic cross section.  If one were to compute
all terms in the series, this would be an exact result.

In fact, a truncated series is sufficient for obtaining the \nnlo{}
cross section to very high numerical precision.  In this case, one
obtains a cancellation of infrared singularities by expanding the other
contributions to the partonic cross section (single real emission, mass
factorization; the dependence of the virtual terms on $x$ is simply
$\propto\delta(1-x)$) to the same order in $(1-x)$ as the double real
emission term.  In \reference{HKggh}, we computed scalar Higgs boson
production to order $(1-x)^{16}$ and arrived at a prediction for the
hadronic cross section that is phenomenologically equivalent to one
based on the closed analytic form of the partonic cross section that has
recently become available~\cite{AnaMel}.  This comes as no great
surprise.  The functions which contribute to the closed analytic form
can all be expanded in terms of $(1-x)$.  For example, the dilogarithm
can be represented as
\begin{equation}
\begin{split}
\Li_2(x) = \frac{\pi^2}{6}
- \sum_{n=1}^{\infty}\frac{(1-x)^n}{n}\left(\frac{1}{n} -
  \ln(1-x)\right)\,.
\end{split}
\end{equation}
Furthermore, steeply falling parton distributions ensure that the
threshold region dominates and that convergence in $(1-x)$ is quite
rapid.

Although the truncated series leads to a physical result that is by
all means equivalent to the exact expression, the approach of
expanding in $(1-x)$ can be taken farther.  If the expansion can be
evaluated up to sufficiently high order in $(1-x)$, one can actually
invert the series and obtain the partonic cross section in closed
analytic form.  This is due to the fact that only a limited number of
functions appear in this closed form representation: logarithms,
dilogarithms, and trilogarithms of various arguments, multiplied by
$1/x$, $1/(1+x)$\break and $(1-x)^{n}, n=-1,0,1,2,3$.  Taking the functions
which appear in the result for the Drell-Yan cross section \cite{DY}
and these possible factors, one finds that carrying out the
expansion to order $(1-x)^{96}$ should suffice to allow inversion of
the series.  In fact, we have carried out the expansion to order
$(1-x)^{100}$ so that we could over-determine the system.  As a check,
this procedure has also been carried out for the scalar Higgs
boson~\cite{kichep02}, and complete agreement with the result of
\reference{AnaMel} was found.

It is interesting to observe that the task of inverting the series is
much simpler when one examines only the difference between the
scalar and pseudo-scalar cases since they are already very similar at
the level of squared amplitudes.  The difference may be inverted with
less than twenty terms.

\section{Partonic results}
As noted above, the partonic cross sections for scalar and pseudo-scalar
Higgs production are very similar, so that many terms cancel in the
difference between the two. Thus, the partonic cross section for
pseudo-scalar Higgs production can be expressed conveniently in terms
of the known scalar Higgs boson cross section (with modified
normalization) plus a remainder. For this purpose, we write
\begin{equation}
\begin{split}
  \hat\sigma_{ab\Phi} &= \sigma^0_\Phi\,\Delta_{ab\Phi}\,,\qquad \Phi \in
  \{H,A\},\qquad a,b \in \{g,q,\bar q\}\,,
\end{split}
\end{equation}
where $\hat\sigma_{ab\Phi}$ is the cross section for the process
$ab \to \Phi+X$. $a$ and $b$ label the partons in the initial state, $\Phi$
means either a scalar ($H$) or pseudo-scalar ($A$) Higgs boson, and $X$
denotes any number of quarks or gluons in the final state.
In the normalization factors, we keep the full top mass dependence:
\begin{equation}
\begin{split}
  \sigma^0_H &= \frac{\pi\,\sqrt{2}\,\gfermi}{256}\left(\api\right)^2\,
  \bigg|\tau_H\left[1+(1-\tau_H)\,f(\tau_H)\right]\bigg|^2\,,\\
  \sigma^0_A &= \frac{\pi\,\sqrt{2}\,\gfermi}{256}
  \left(\api\right)^2\,\Big|g_t\,\tau_A\,f(\tau_A)\Big|^2\,,
  \qquad\qquad \tau_\Phi = \frac{4M_t^2}{M_\Phi^2}\,,
  \label{eq::1loop}
\end{split}
\end{equation}
where $\gfermi \approx 1.664\times 10^{-5}\,{\rm GeV}^{-2}$, and $g_t$
has been introduced in \eqn{eq::ttA}. The one-loop function appearing in
\eqn{eq::1loop} is defined by
\begin{equation}
\begin{split}
f(\tau) &= \left\{
  \begin{array}{ll}
    \arcsin^2\frac{1}{\sqrt{\tau}}\,, & \tau\geq 1\,,\\
    -\frac{1}{4}\left[
      \ln\frac{1+\sqrt{1-\tau}}{1-\sqrt{1-\tau}} -i\pi\right]\,, 
    & \tau < 1\,.
  \end{array}
  \right.
\end{split}
\end{equation}
For completeness, we note that in the heavy-top limit, the
normalization factors approach the values
\begin{equation}
\begin{split}
\sigma_H^0\ &\stackrel{M_t\to \infty}{\longrightarrow}
\ \frac{\pi\,\sqrt{2}\,\gfermi}{576}\,,\qquad\quad
\sigma_A^0\ \stackrel{M_t\to \infty}{\longrightarrow}
\ \frac{\pi\,\sqrt{2}\,\gfermi}{256}\,|g_t|^2\,.
\end{split}
\end{equation}
The kinematic terms are written as a perturbative expansion:
\begin{equation}
\begin{split}
\Delta_{ab\Phi}(x) &= \delta_{ag}\delta_{bg}\,\delta(1-x)
+ \api\,\Delta_{ab\Phi}^{(1)}(x) +
\left(\api\right)^2\,\Delta_{ab\Phi}^{(2)}(x)
+ \order{\alpha_s^3}\,.
\end{split}
\end{equation}

The results for scalar Higgs production up to \nnlo{} ($\Phi=H$)
can be found in {\rm Refs.}\,\cite{soft,HKggh,AnaMel}.

With the help of these expressions, the corresponding results for the
pseudo-scalar Higgs production cross section can be written in a rather
compact form. 
At \nlo{}, the result has been evaluated some time ago~\cite{nlo,nloeff}:
\begin{equation}
\begin{split}
\Delta_{abA}^{(1)}(x) &= \Delta_{abH}^{(1)}(x) + 
\frac{1}{2}\,\delta_{ag}\delta_{bg}\,\delta(1-x)\,,
\qquad a,b\in\{g,q,\bar q\}\,.
\label{eq::diffah1}
\end{split}
\end{equation}

The main result of this paper are the \nnlo{} terms in $\Delta_{abA}$.
We find:
\begin{equation}
\begin{split}
\Delta_{ggA}^{(2)}&(x) = \Delta_{ggH}^{(2)}(x) +
\bigg[ \frac{1939}{144} - \frac{19}{8}\,\lht + 3\,\zeta_2
\bigg]\,\delta(1-x)
+ 6\,{\cal D}_1(x)
\\&
- (12\,x - 6\,x^2 + 6\,x^3)\,\ln(1-x)
- 9\,x\,\ln^2(x)
\\&
+ \frac{3}{2}\,\frac{(10 - x - 13\,x^2 + 4\,x^3 - 2\,x^4)}{1 - x}\,\ln(x)
+ \frac{(154 - 189\,x + 24\,x^2 + 11\,x^3)}{4}
\\&
+ n_f\bigg[
\left( -\frac{13}{16} - \frac{2}{3}\,\lht + 2\,\delta_2 \right)\,\delta(1-x)
+ \frac{2}{3}\,x\,\ln^2(x)
+ x\,\ln(x)
\\&\qquad
- \frac{(1 - 11\,x + 10\,x^2 )}{6}
\bigg]\,,\quad
 \mbox{with}\qquad
\delta_2 = -\frac{1}{4} + \frac{1}{2}\,\lht\,,\\
\Delta_{gqA}^{(2)}&(x) = \Delta_{gqH}^{(2)}(x)
+ \frac{(4 - 4\,x + 2\,x^2)}{3}\ln(1-x)
-\frac{28}{9}\,x\,\ln^2(x)
\\&
+ \frac{(22 + 30\,x - x^2)}{3}\ln(x)
+ \frac{(337 - 382\,x + 51\,x^2)}{18}\,,
\\
\Delta_{qqA}^{(2)}&(x) = \Delta_{qqH}^{(2)}(x)
-\frac{64}{27}\,x\,\ln^2(x)
+ \frac{16}{27}\,( 6 + 11\,x )\,\ln(x)
+ \frac{8}{27}\,(37 - 40\,x + 3\,x^2)\,,
\\
\Delta_{q\bar{q}A}^{(2)}&(x) = \Delta_{q\bar{q}H}^{(2)}(x) +
\frac{32}{27}\,x\,\ln^2(x)
+ \frac{32}{27}\,( 3 + 8\,x)\,\ln(x)
\\&
+ \frac{16}{27}\,(11 - x - 9\,x^2 - x^3)
+ n_f\,\bigg[
-\frac{32}{27}\,x\,\ln(x)
- \frac{16}{27}\,(1-x^2)
\bigg]\,,
\\
\Delta_{qq'A}^{(2)}&(x) = \Delta_{qq'H}^{(2)}(x)
-\frac{16}{9}\,x\,\ln^2(x)
+ \frac{16}{9}\,(2 + 3\,x)\,\ln(x)
+ \frac{8}{9}\,( 11 - 12\,x + x^2 )\,,
\label{eq::diffah2}
\end{split}
\end{equation}
where $\lht \equiv \ln(\mahiggs^2/M_t^2)$, $\zeta_n\equiv\zeta(n)$ is
Riemann's $\zeta$ function ($\zeta_2 = \pi^2/6 = 1.64483\ldots$,
$\zeta_3 = 1.20206\ldots$) and ${\cal D}_{n}(x) \equiv
[\ln^{n}(1-x)/(1-x)]_+$.  Of course one must change $\mhiggs\to\mahiggs$
wherever it appears in $\Delta_{abH}$.  For the sake of completeness,
we list the full result for $\Delta_{abA}$ in App.\,\ref{app::full}.
The difference $\Delta_{abA} - \Delta_{abH}$ can be expanded readily
in terms of $(1-x)$ in order to bring it to a form consistent with the
results of \reference{HKggh}.

The only contribution that originates from the presence of the
(renormalized!) operator $\opo_2$ is the term $\delta_2$ in the
equations above. We obtain it by computing the interference of
diagram~$(a)$ in \fig{fig::o1} with diagram~$(a)$ of \fig{fig::o2}.
Alternatively, it can be derived from \eqn{eq::o1o2rel} by using the
\lo{} result for $\Delta_{ggA}$:
\begin{equation}
\begin{split}
  (\coeff_1\coeff_2)^{-1}\,n_f\,\delta_2\,\left(\api\right)^4\,\delta(1-x)
  &=
  \api\,\frac{n_f}{8}\left[\coeff_1^{-2}\left(\api\right)^2\,\delta(1-x)
  \right]\,,
\end{split}
\end{equation}
which leads to the same result for $\delta_2$ as above. This provides a
welcome check on the normalization of the contribution from
$\opo_2$.

\section{Hadronic results}
In complete analogy to scalar Higgs production, one has to convolute the
partonic rate with the proper parton distribution functions, in order to
arrive at a physical prediction for the hadronic production rate:
\begin{equation}
\begin{split}
\sigma_{h_1h_2}(s) &= \sum_{a,b}\,\int_0^1\!d\tau\!\int_\tau^1
   \frac{dx_a}{x_a}\left[f_{a/h_1}(x_a)\,f_{b/h_2}(\tau/x_a)\right]\,
   {\hat{\sigma}}_{ab}(\hat{s}=s\tau).
\end{split}
\end{equation}
A consistent \nnlo{} hadronic result requires not only a \nnlo{}
partonic cross section, but also \pdf{}s that have been evaluated at
\nnlo{}. Strictly speaking, such a set of \pdf{}s is not yet
available, because the \nnlo{} evolution equation is not fully
known. Nevertheless, we use an approximate set of
\pdf{}s~\cite{mrstnnlo}, based on approximations to the evolution
equation~\cite{evolve} derived from the available moments of the
structure functions~\cite{structs}.  At lower order, we use the
corresponding lower order \pdf{}s~\cite{MRST2001}.

\FIGURE{
  \leavevmode
  \begin{tabular}{cc}
    \epsfxsize=17em
    \epsffile[110 265 465 560]{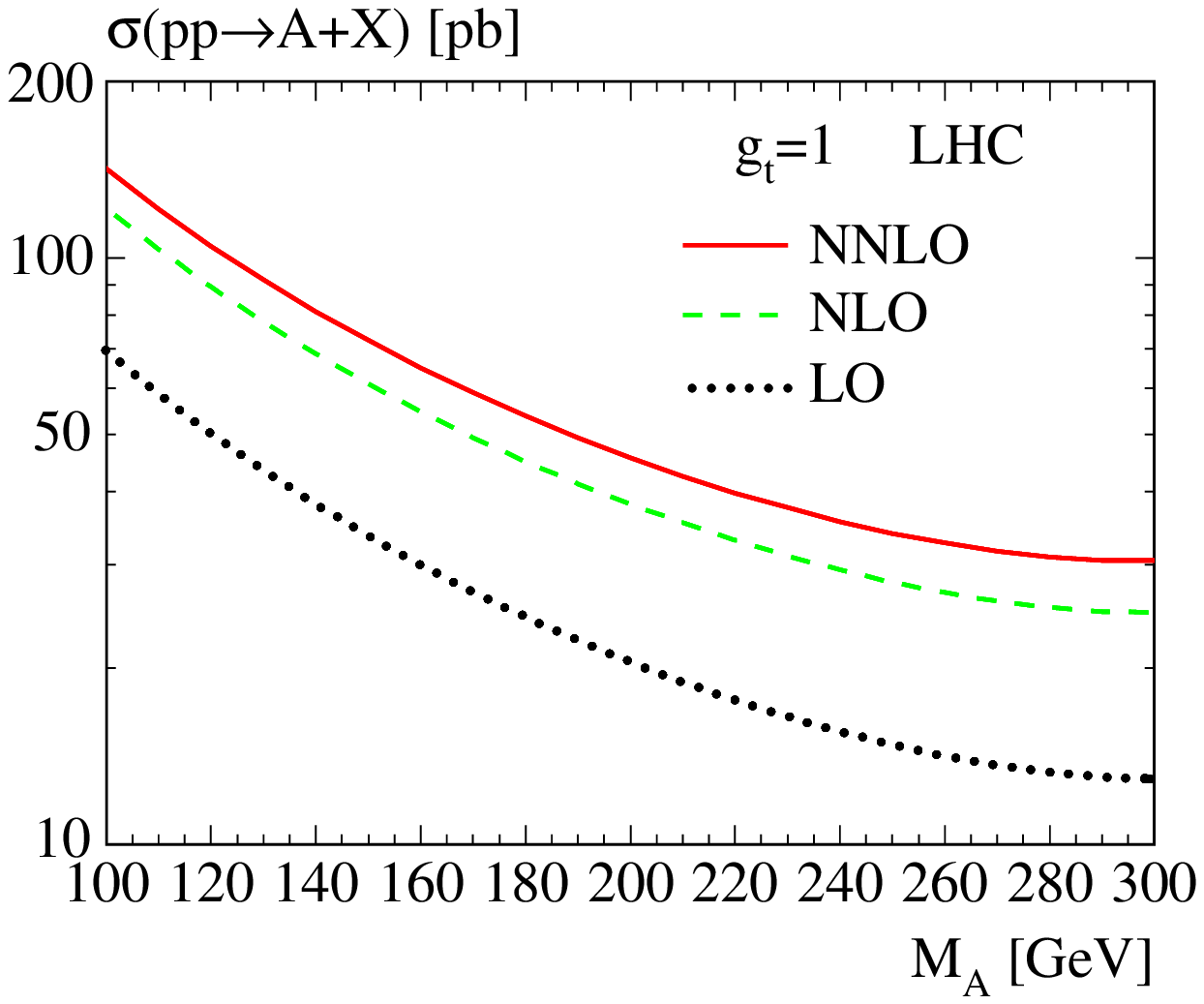} &
    \epsfxsize=17em
      \epsffile[110 265 465 560]{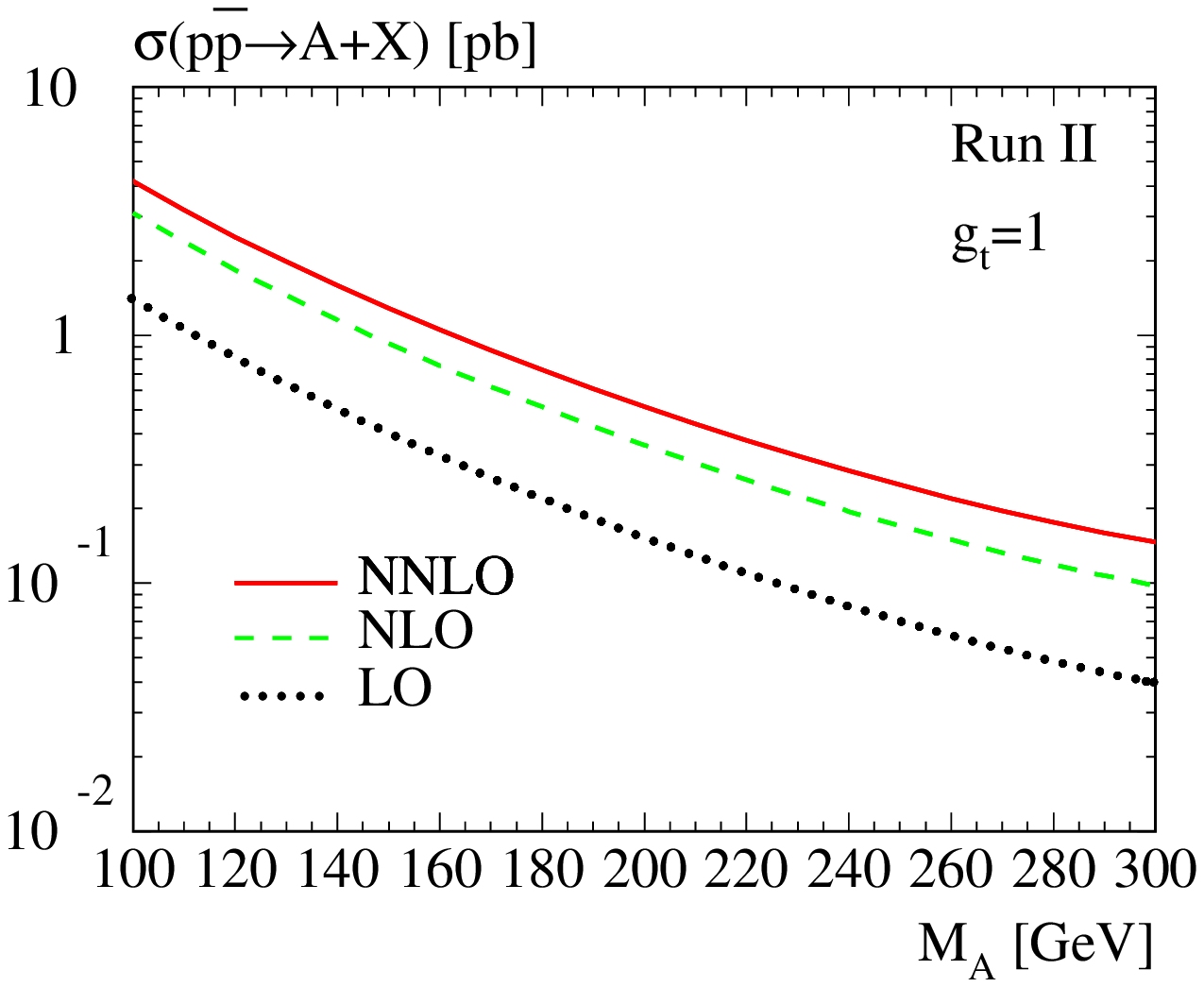}\\
      $(a)$ & $(b)$
  \end{tabular}
  \caption[]{\label{fig::sig14}\sloppy
    Total cross section for inclusive production of pseudo-scalar Higgs
    bosons ($A$) at $(a)$ the \lhc{} ($\sqrt{s}=14$\,TeV) and $(b)$
    Tevatron Run\,II ($\sqrt{s}=2$\,TeV).  The coupling constant for the
    coupling of $A$ to top quarks is such that $g_t=1$. The cross section for
    other values of $g_t$ (e.g., $g_t=\cot\beta$ in the \mssm{}) can be
    obtained by scaling the curves with $|g_t|^2$.  } }
\FIGURE{
  \leavevmode
  \begin{tabular}{cc}
    \epsfxsize=25em
      \epsffile[110 265 465 560]{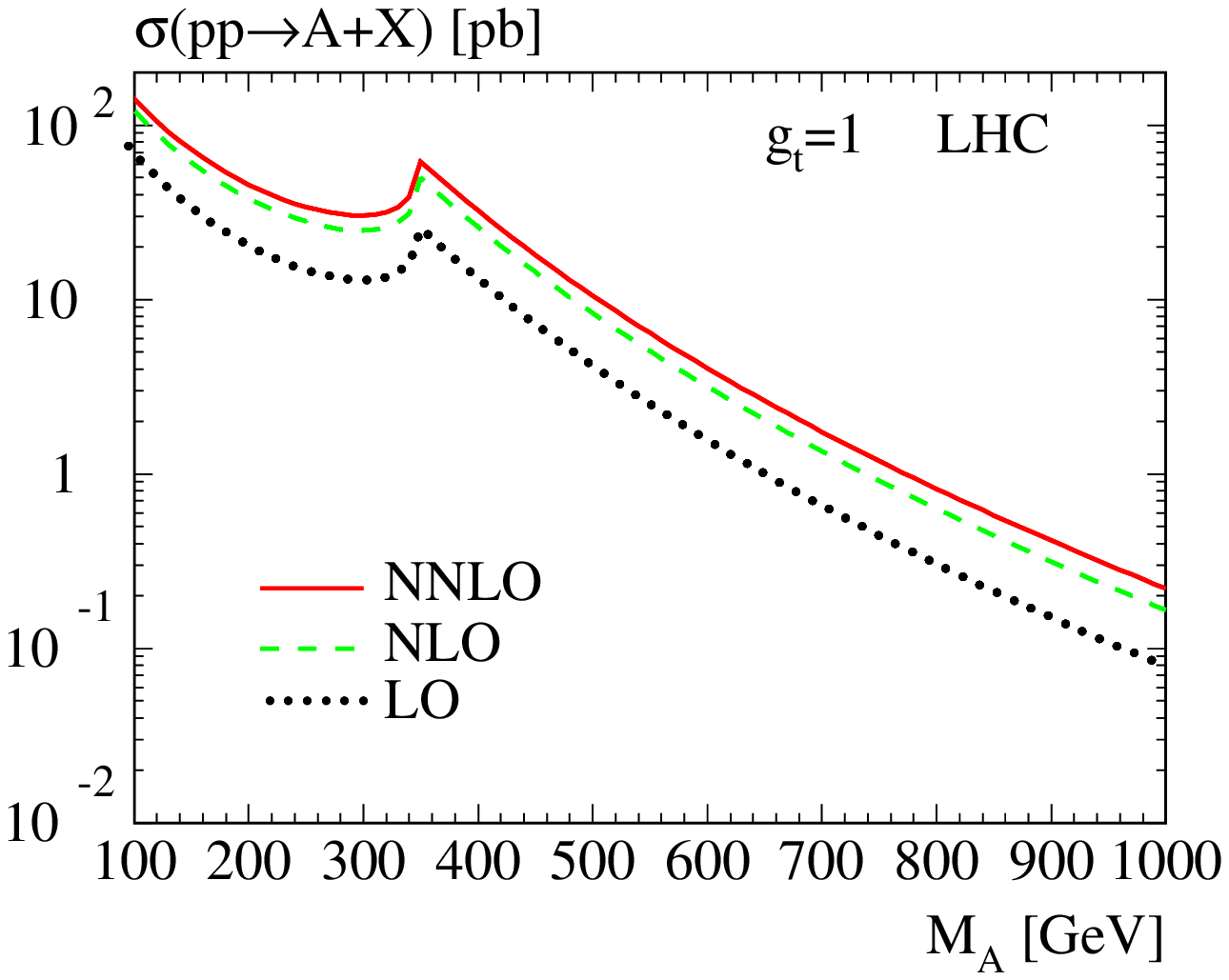}
  \end{tabular}
  \caption[]{\label{fig::sig14big}\sloppy
    Total cross section for inclusive production of pseudo-scalar Higgs
    bosons ($A$) at the \lhc{} ($\sqrt{s}=14$\,TeV), as in
    \fig{fig::sig14} for a larger mass range.
    The cusp is an effect of the top-quark threshold. } 
  }
\FIGURE{
  \leavevmode
  \begin{tabular}{ccc}
    \mbox{\hspace{2em}}
    \epsfxsize=11em
    \epsffile[184 210 411 610]{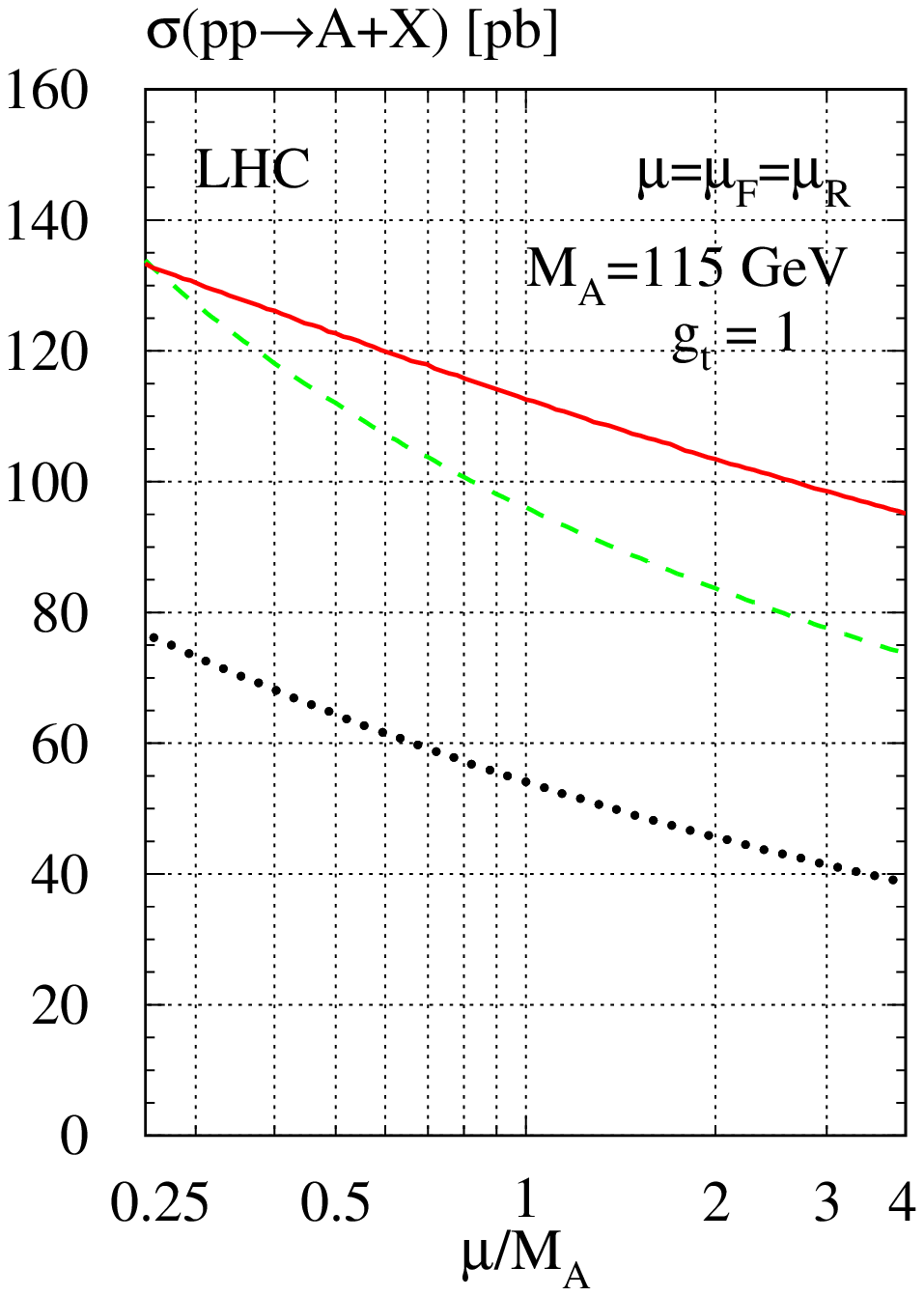} &
    \epsfxsize=11em
    \epsffile[184 210 411 610]{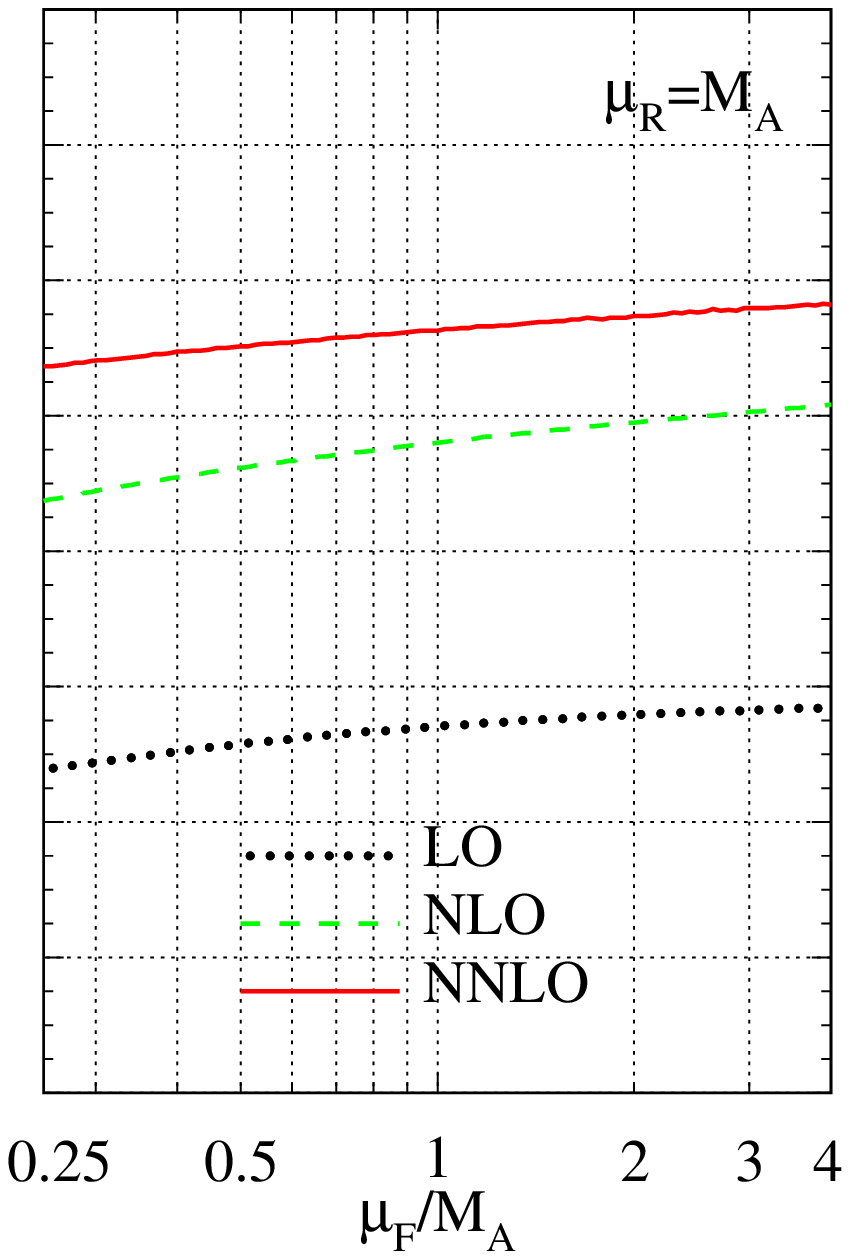} &
    \epsfxsize=11em
    \epsffile[184 210 411 610]{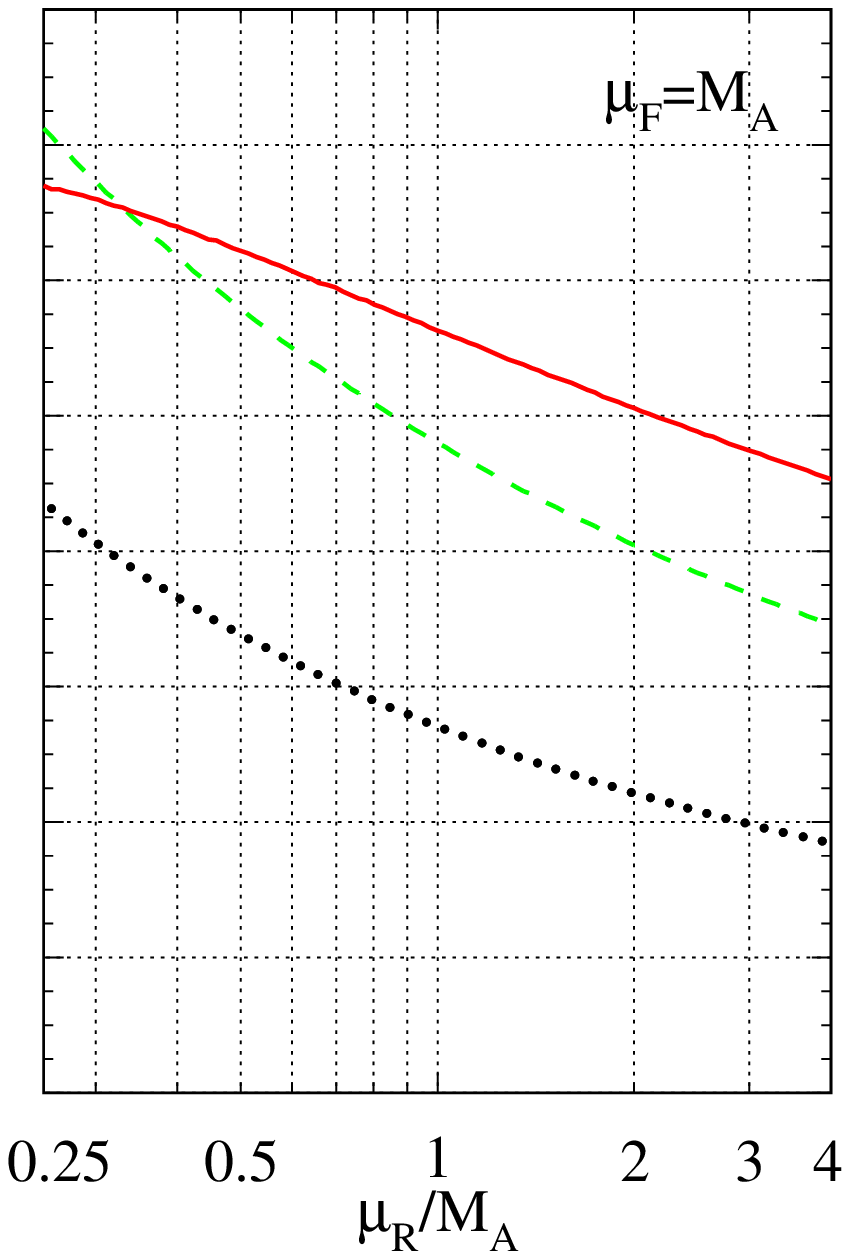}\\
    $(a)$ & $(b)$ & $(c)$
  \end{tabular}
  \caption[]{\label{fig::mu14}\sloppy
    Scale dependence of the cross section for an
    $\mahiggs=115$ GeV pseudo-scalar Higgs boson at the \lhc{}
    ($\sqrt{s}=14$\,TeV) $(a)$ varying $\mu_F = \mu_R$, $(b)$
    varying $\mu_F$, $\mu_R = \mahiggs$ and $(c)$ varying $\mu_R$,
    $\mu_F = \mahiggs$. 
    }
  }

\FIGURE{
  \leavevmode
  \begin{tabular}{ccc}
    \mbox{\hspace{2em}}
    \epsfxsize=11em
    \epsffile[184 210 411 610]{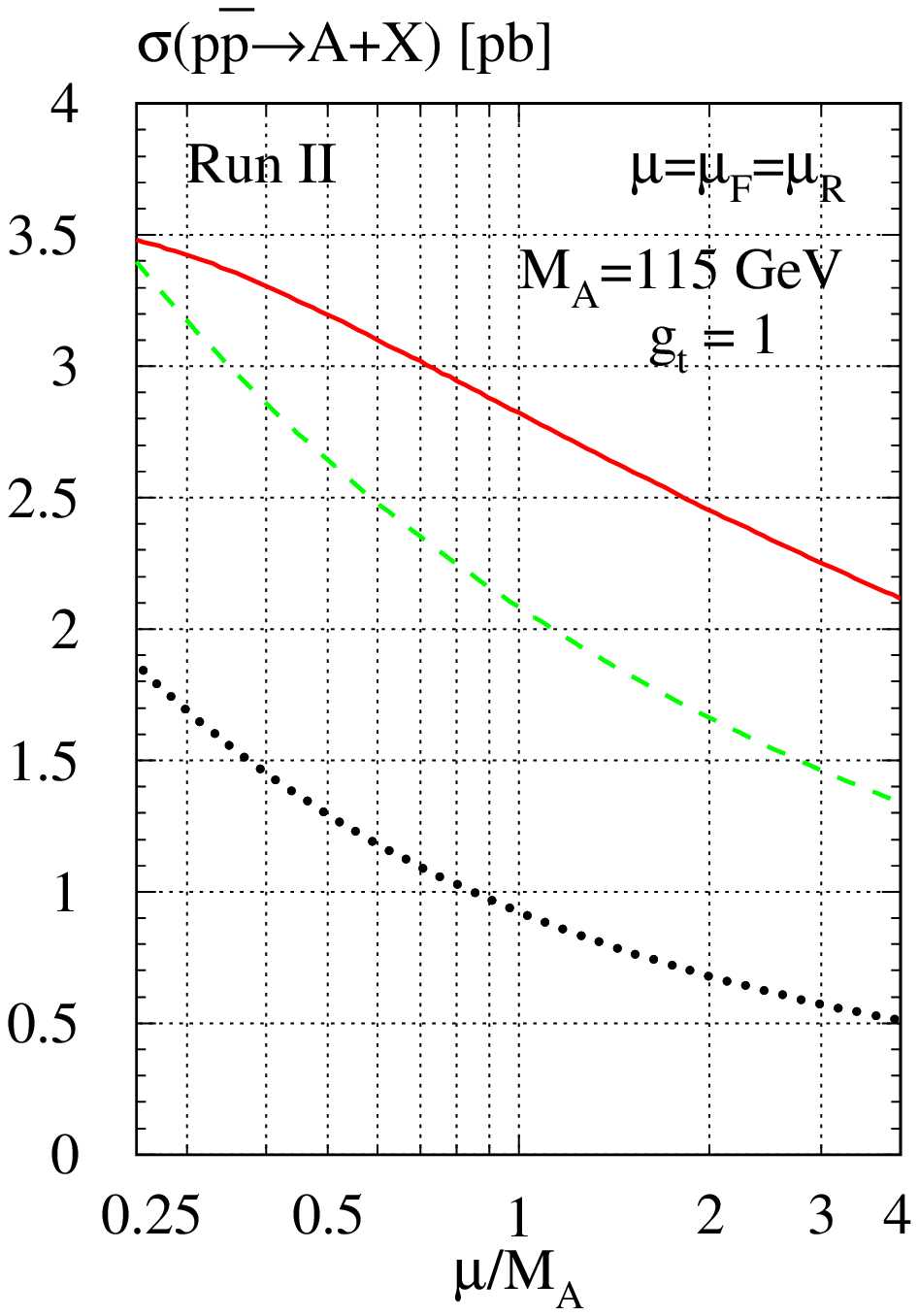} &
    \epsfxsize=11em
    \epsffile[184 210 411 610]{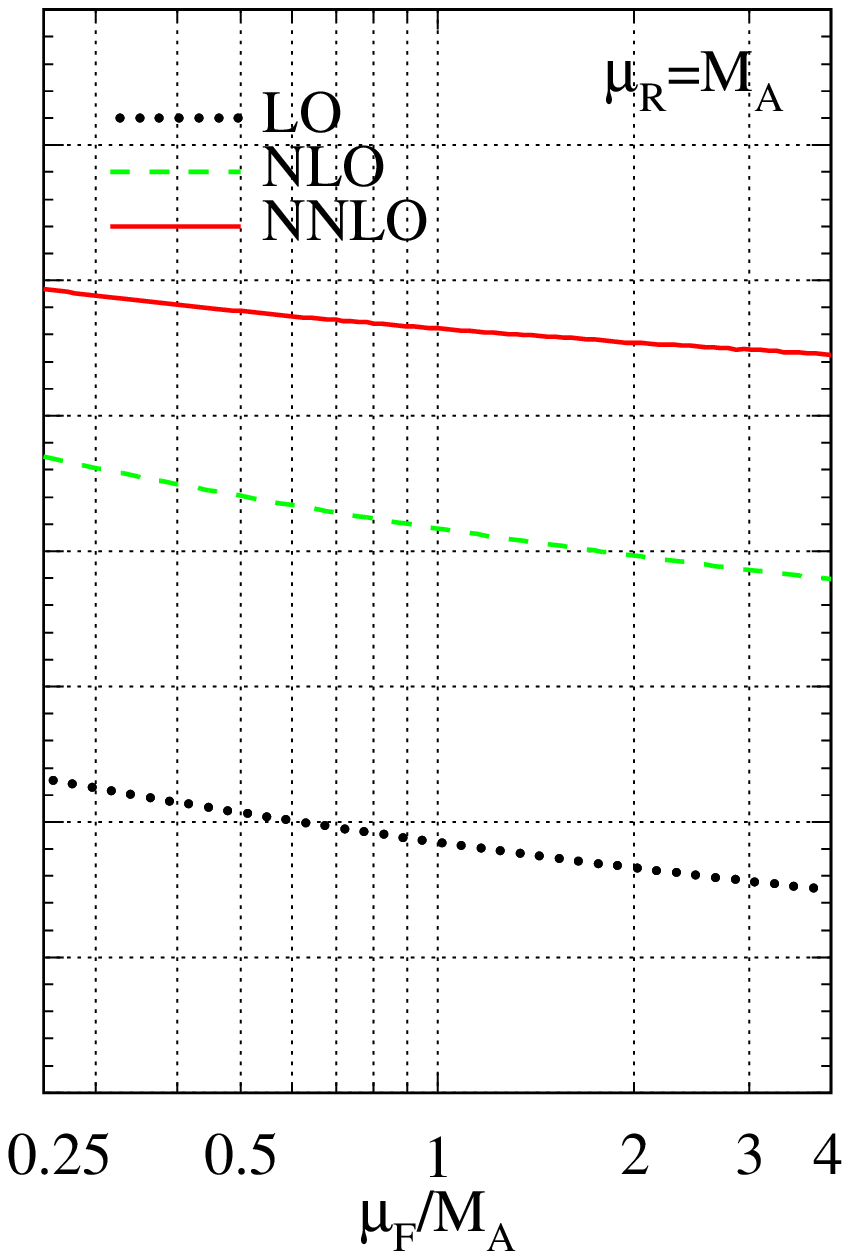} &
    \epsfxsize=11em
    \epsffile[184 210 411 610]{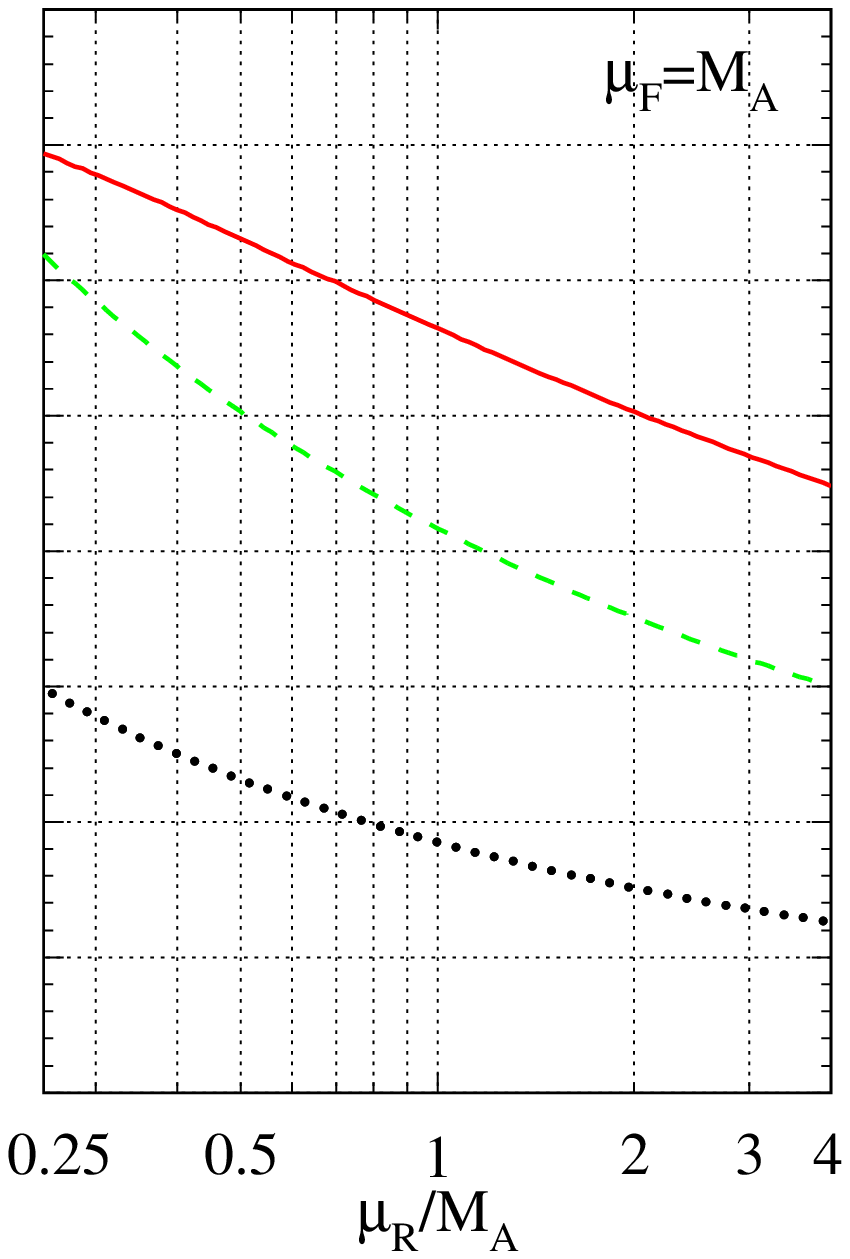}\\
    $(a)$ & $(b)$ & $(c)$
  \end{tabular}
  \caption[]{\label{fig::mu2}\sloppy
    Same as \fig{fig::mu14} for the Tevatron Run\,II ($\sqrt{s}=2$\,TeV).
    }
  }

\fig{fig::sig14} shows the total cross section as a function of the
pseudo-scalar Higgs boson mass, $\mahiggs$, at $(a)$ the \lhc{}, and
$(b)$ Tevatron Run\,II. \fig{fig::sig14big} shows the curves for the
\lhc{} over a larger mass range. The renormalization and factorization
scales have been set to $\mahiggs$.
The cusp in \fig{fig::sig14big} is a leading order effect caused by the $t\bar
t$ threshold, cf.\,\eqn{eq::1loop}.  Since we do not want to specify
a particular extension of the Standard Model, we choose the coupling of
the pseudo-scalar Higgs to the top quarks such that $g_t=1$
(cf.\,\eqn{eq::ttA}). The total cross section scales with $|g_t|^2$, so
that the actual numbers can be easily obtained from the ones shown in
the figure. As already noted, the behavior of the corrections is very
similar to the ones for scalar Higgs production: The \nnlo{} corrections
are significantly smaller than the \nlo{} corrections, indicating a
nicely converging result with uncertainties well under control.

This is further affirmed in Figs.\,\ref{fig::mu14} and \ref{fig::mu2},
which show the variation of the cross section with the renormalization
and factorization scales for a fixed Higgs mass $\mahiggs=115$\,GeV.
The scales are varied between $\mahiggs/4$ and $4\mahiggs$: in
sub-panels $(a)$, the renormalization scale and the factorization
scale are identified and varied simultaneously; in sub-panels $(b)$,
the renormalization scale $\mu_R$ is identified with $\mahiggs$, and
$\mu_F$ is varied, while in sub-panels $(c)$, $\mu_F=\mahiggs$ is
fixed and $\mu_R$ is varied. One observes a clear reduction of the
scale dependence at \nnlo{} with respect to \nlo{}. Compared to the
\lo{} curve, there is a significant {\it relative} improvement in the
scale variation, while the absolute ranges of variation at \lo{} and
\nnlo{} are comparable. It is also clear that the scale dependence is
dominated by renormalization scale dependence; factorization scale
dependence is quite small.  Using the rather conservative range of
$\mahiggs/4<\mu<4\mahiggs$, one arrives at an uncertainty in $\sigma$
from scale variation of about $\pm 30\%$ ($\pm 65\%$) at \lo{}, $\pm
30\%$ ($\pm 50\%$) \nlo{}, and less than $\pm 20\%$ ($\pm 25\%$) at
\nnlo{} for the \lhc{} (numbers in brackets for Tevatron Run\,II).
Varying $\mu$ between $\mahiggs/2$ and $2\mahiggs$ results in a
variation of $\sigma$ of $\pm 20\%$ ($\pm 40\%$) at \lo{}, $\pm 15\%$
($\pm 25\%$) at \nlo{}, and less than $\pm 10\%$ ($\pm 15\%$) at
\nnlo{}.

\section{Conclusions}
The hadronic cross section for the production of a pseudo-scalar Higgs
boson has been calculated at \nnlo{} in {\abbrev QCD}. We find that
the corrections are similar to the scalar case, both in their
magnitude and in their uncertainty due to scale dependence.  While
this uncertainty is still rather large, it seems that the \nnlo{}
calculation yields a reliable prediction for the total rate.

\paragraph{Note Added:}
As we completed this manuscript, we became aware of a similar paper by
Anastasiou and Melnikov\cite{AnaMel2}.  We have compared results for
the partonic cross sections and find complete agreement.

\paragraph{Acknowledgments.}
R.V.H.\ thanks the High Energy Theory group at Brookhaven National
Laboratory, where part of this work has been done, for hospitality.
The work of W.B.K.\ was supported by the U.S.\,Department of Energy
under Contract No.~{\abbrev DE-AC02-98CH10886}.

\begin{appendix}

\section{Analytic results}\label{app::full}
In this appendix, the analytic expressions for the partonic cross
section of pseudo-scalar Higgs production at \nnlo{} are listed. Note
that using the formulas of Eqs.\,(\ref{eq::diffah1}) and
(\ref{eq::diffah2}), one can transform these expressions into the
corresponding ones for scalar Higgs production.

At \nlo{}~\cite{nlo,nloeff}, the result for the gluon-gluon sub-process is
\begin{equation}
\begin{split}
\Delta_{ggA}^{(1)}(x)& = 
(6+6\,\zeta_2)\,\delta(1-x) + 12\,{\cal D}_1(x)
- (24\,x - 12\,x^2 + 12\,x^3)\,\ln(1-x)
\\&
- 6\frac{(1 - x + x^2)^2}{1-x}\ln(x)
- \frac{11}{2}(1-x)^3\,.
\end{split}
\end{equation}
For the quark-gluon channel one finds
\begin{equation}
\begin{split}
\Delta_{gqA}^{(1)}(x)& = 
\frac{(8 - 8\,x + 4\,x^2)}{3}\ln(1-x)
-\frac{(4 - 4\,x + 2\,x^2)}{3}\ln(x)
-\frac{(3 - 6\,x + x^2)}{3}
\,,
\end{split}
\end{equation}
and for the quark--anti-quark channel
\begin{equation}
\begin{split}
\Delta_{q\bar qA}^{(1)}(x) = &
\frac{32}{27}(1-x)^3\,.
\end{split}
\end{equation}

The contributions from different sub-processes at \nnlo{} are written as
\begin{equation}
\begin{split}
\Delta_{abA}^{(2)} &= \Delta_{abA}^{A} + n_f\,\Delta_{abA}^{F}\,,
\end{split}
\end{equation}
where $n_f$ is the number of light (i.e., massless in our approach)
quark flavors.

For the sub-process with two gluons in the initial state, we find

{
\begin{equation}
\begin{split}
&\Delta_{ggA}^{A}(x)=
\left[
 \frac{741}{8} 
+ \frac{139}{2}\zeta_2 
- \frac{165}{4}\zeta_3
- \frac{9}{20}\zeta_2^2 
\right]\delta(1-x)
\\&
- \left[
\frac{101}{3 }
- 33\,\zeta_2 
- \frac{351}{2}\,\zeta_3
\right]{\cal D}_0(x)
+ \left[
139 
- 90\,\zeta_2
\right]{\cal D}_1(x)
-33\, {\cal D}_2(x)
+ 72\,{\cal D}_3(x)
\\&
-(144\,x - 72\,x^2 + 72\,x^3)
 \ln^3(1-x)
-(297 - 381\,x + 348\,x^2 - 330\,x^3)
 \ln^2(1-x)
\\&
-\frac{9}{2}\,\frac{(31 - 30\,x + 93\,x^2 - 94\,x^3 + 31\,x^4)}{1-x}
 \ln^2(1-x)\ln(x)
\\&
+\left[
 \frac{(2027 - 2735\,x + 2182\,x^2 - 2583\,x^3)}{4}
+(180\,x - 90\,x^2 + 90\,x^3)\zeta_2
 \right]\ln(1-x)
\\&
+3\,\frac{(88 - 211\,x + 312\,x^2 - 365\,x^3 + 187\,x^4)}{1-x}
 \ln(1-x)\ln(x)
\\&
+9\,\frac{(7 + 3\,x + 19\,x^2 - 3\,x^3 - 19\,x^4 + 9\,x^5)}{1-x^2}
 \ln(1-x)\ln^2(x)
\\&
+36\frac{(1 - 6\,x - 13\,x^2 - 6\,x^3 + x^4)}{1+x}
 \ln(1-x)\Li_2(1-x)
\\&
-18\frac{(1 + 2\,x + 3\,x^2 + 2\,x^3 + x^4)}{1+x}
 \ln(1-x)\Li_2(1-x^2)
\\&
-\frac{9}{2}\frac{(24 - 38\,x + 8\,x^2 + 54\,x^3 - 19\,x^4 + 9\,x^5)}{1 - x^2}
 \Li_3(1-x)
\\&
-\frac{9}{2}\frac{(27 + 35\,x + 75\,x^2 - 29\,x^3 - 78\,x^4 + 6\,x^5)}{1-x^2}
 \Li_3\left(-\frac{(1-x)}{x}\right)
\\&
+\frac{9}{8}\frac{(1 + 2\,x + 3\,x^2 + 24\,x^3 + 16\,x^4)}{1 + x}
 \Li_3(1-x^2)
\\&
-\frac{9}{8}\frac{(1 + 2\,x + 3\,x^2 - 8\,x^3 - 8\,x^4)}{1 + x}
 \Li_3\left(-\frac{(1-x^2)}{x^2}\right)
\\&
-\frac{9}{2}\frac{(7 + 14\,x + 21\,x^2 + 8\,x^3 + 4\,x^4)}{1+x}
 \left[\Li_3\left(\frac{1-x}{1+x}\right)
  -\Li_3\left(-\frac{1-x}{1+x}\right)\right]
\\&
-\frac{3}{4}\frac{(317 - 398\,x - 87\,x^2 + 300\,x^3 - 121\,x^4)}{1-x}
 \Li_2(1-x)
\\&
+\frac{9}{2}\frac{(11 + 31\,x + 59\,x^2 - 25\,x^3 - 65\,x^4 + 11\,x^5)}{1-x^2}
 \Li_2(1-x)\ln(x)
\\&
+\frac{(42 + 36\,x - 63\,x^2 - 33\,x^3)}{4}
 \Li_2(1-x^2)
+\frac{9}{4}\frac{(5 + 10\,x + 15\,x^2 - 2\,x^4)}{1+x}
 \Li_2(1-x^2)\ln(x)
\\&
+\frac{3}{4}\frac{(21 + 23\,x + 41\,x^2 - 37\,x^3 - 44\,x^4 + 4\,x^5)}{1 - x^2}
 \ln^3(x)
\\&
-\frac{3}{8}\frac{(154 - 365\,x + 675\,x^2 - 827\,x^3 + 374\,x^4)}{1 - x}
 \ln^2(x)
\\&
-\frac{1}{8}\frac{(2213 - 5599\,x + 6603\,x^2 - 7003\,x^3 + 4342\,x^4)}{1 - x}
 \ln(x)
\\&
+9\,\frac{(9 - 2\,x + 27\,x^2 - 34\,x^3 + 9\,x^4)}{1 - x}
 \zeta_2\,\ln(x)
-\frac{(16309 - 20611\,x + 23819\,x^2 - 22749\,x^3)}{48}
\\&
+\frac{3}{4}\,(319 - 277\,x + 233\,x^2 - 363\,x^3) \zeta_2
-\frac{351}{2}\,\,(2\,x - x^2 + x^3)
 \zeta_3
\end{split}
\end{equation}
}
%
%
and 
\begin{equation}
\begin{split}
&\Delta_{ggA}^{F}(x)=
\left[-\frac{689}{72}
+ \lht
- \frac{5}{3}\,\zeta_2 
+ \frac{5}{6}\,\zeta_3\right]\,\delta(1-x)
\\&
+\left[\frac{14}{9} - 2\,\zeta_2\right]\,{\cal D}_0(x)
-\frac{10}{3}\, {\cal D}_1(x)
+2 \, {\cal D}_2(x)
\\&
+\frac{2}{9}(8 - 12\,x + 3\,x^2 - 17\,x^3)
 \ln^2(1-x)
+\frac{8}{3}(x + x^2)
 \ln^2(1-x)\ln(x)
\\&
-\frac{(922 - 294\,x + 249\,x^2 - 1570\,x^3)}{108}
 \ln(1-x)
\\&
-\frac{2}{9}\frac{(17 + 7\,x + 21\,x^2 - 61\,x^3 + 25\,x^4)}{1-x}
 \ln(1-x)\ln(x)
\\&
-\frac{8}{3}(x + x^2)\ln^2(x)\ln(1-x)
+\frac{16}{3}(x + x^2) \ln(1-x)\,\Li_2(1-x)
\\&
-\frac{(2 + 14\,x + 17\,x^2)}{6} \Li_3(1-x)
-\frac{(2 - 34\,x - 31\,x^2)}{12} \Li_3\left(-\frac{(1-x)}{x}\right)\,.
\\&
+\frac{1}{36}\frac{(68 - 302\,x + 21\,x^2 + 227\,x^3 + 4\,x^4)}{1-x}
 \Li_2(1-x)
\\&
+\frac{(2 - 50\,x - 47\,x^2)}{12}
 \Li_2(1-x)\ln(x)
\\&
+\frac{(2 + 6\,x + 9\,x^2)}{72}
 \ln^3(x)
+\frac{1}{72}\frac{(68 + 100\,x + 69\,x^2 - 351\,x^3 + 132\,x^4)}{1 - x}
 \ln^2(x)
\\&
+\frac{1}{216}\frac{(1282 - 382\,x + 117\,x^2 - 3041\,x^3 + 2384\,x^4)}{1 - x}
 \ln(x)
-\frac{8}{3}(x + x^2)
 \zeta_2\,\ln(x)
\\&
+\frac{(12707 - 606\,x + 1641\,x^2 - 17774\,x^3)}{1296}
-\frac{2}{9}(8 - 12\,x + 3\,x^2 - 17\,x^3)\,
 \zeta_2
\end{split}
\end{equation}

For the quark-gluon channel, we have:
\begin{equation}
\begin{split}
&\Delta_{gqA}^{A}(x)=
\frac{367}{54}(2 - 2\,x + x^2)
 \ln^3(1-x)
\\&
-\frac{(2592 - 2278\,x - 111\,x^2 - 288\,x^3)}{36}
 \ln^2(1-x)
-\frac{(642 + 190\,x + 553\,x^2)}{18}
 \ln^2(1-x)\ln(x)
\\&
+\left[
\frac{(23887 - 17388\,x - 2538\,x^2 - 784\,x^3)}{162}
-\frac{50}{9}(2 - 2\,x + x^2)
 \zeta_2\,
\right] \ln(1-x)
\\&
+\frac{(1665 - 2040\,x + 174\,x^2 - 400\,x^3)}{27}
 \ln(1-x)\ln(x)\!
+\frac{4(38 + 21\,x + 39\,x^2)}{9}
 \ln(1-x)\ln^2(x)
\\&
-\frac{2}{9}(46 + 298\,x + 139\,x^2)
 \ln(1-x)\,\Li_2(1-x)
-2(2 + 2\,x + x^2)
 \ln(1-x)\,\Li_2(1-x^2)
\\&
-\frac{2}{9}(42 - 142\,x - x^2)
 \Li_3(1-x)
-\frac{(302 + 474\,x + 339\,x^2)}{9}
 \Li_3\left(-\frac{(1-x)}{x}\right)
\\&
-\frac{(2 + 2\,x + x^2)}{2}
 \Li_3(1-x^2)
-\frac{(2 + 2\,x + x^2)}{2}
 \Li_3\left(-\frac{(1-x^2)}{x^2}\right)
\\&
-4(2 + 2\,x + x^2)
\left[\Li_3\left(\frac{1-x}{1+x}\right)
-\Li_3\left(-\frac{1-x}{1+x}\right)\right]
\\&
-\frac{(979 - 144\,x - 215\,x^2 + 52\,x^3)}{18}
 \Li_2(1-x)
+\frac{(142 + 374\,x + 245\,x^2)}{9}
 \Li_2(1-x)\ln(x)
\\&
+\frac{(166 + 222\,x + 33\,x^2 + 4\,x^3)}{54}
 \Li_2(1-x^2)
+2(2 + 2\,x + x^2)
 \ln(x)\,\Li_2(1-x^2)
\\&
+\frac{(133 + 202\,x + 115\,x^2)}{27}
 \ln^3(x)
-\frac{(837 - 1296\,x + 234\,x^2 - 226\,x^3)}{54}
 \ln^2(x)
\\&
-\left[
 \frac{(22042 - 32040\,x - 5847\,x^2 - 2464\,x^3)}{324}
-\frac{(194 + 222\,x + 213\,x^2)}{9} \zeta_2
 \right]\ln(x)
\\&
-\frac{(173719 - 156324\,x - 12687\,x^2 - 6148\,x^3)}{1944}
+\frac{(1071 - 710\,x - 130\,x^2 - 144\,x^3)}{18}\,\zeta_2
\\&
+\frac{311}{18}(2 - 2\,x + x^2)
\, \zeta_3
\end{split}
\end{equation}
%
%
and
\begin{equation}
\begin{split}
&\Delta_{gqA}^{F}(x)=
\frac{(2 - 2\,x + x^2)}{18}
 \ln^2(1-x)
\\&
-\frac{(13 - 16\,x + 9\,x^2)}{9}
 \ln(1-x)
-\frac{(4 - 4\,x + 2\,x^2)}{9}
 \ln(1-x)\ln(x)
\\&
+\frac{(2 - 2\,x + x^2)}{9}
 \ln^2(x)
+\frac{(29 - 38\,x + 19\,x^2)}{27}
 \ln(x)
+\frac{(265 - 418\,x + 179\,x^2)}{162}
\,.
\end{split}
\end{equation}


For the scattering of two identical quarks, we find
\begin{equation}
\begin{split}
\Delta_{qqA}^{A}&(x)=
-\frac{32}{9} (3 - 2\,x - x^2)
 \ln^2(1-x)
-\frac{16}{9} (4 + 4\,x + x^2)
 \ln^2(1-x)\ln(x)
\\&
+\frac{4}{3} (17 - 12\,x - 5\,x^2)
 \ln(1-x)
+\frac{8}{9} (12 - 8\,x - 5\,x^2)
 \ln(1-x)\ln(x)
\\&
+\frac{8}{9}(4 + 4\,x + x^2)
 \ln(1-x)\ln^2(x)
-\frac{32}{9} (4 + 4\,x + x^2)
 \ln(1-x)\Li_2(1-x)
\\&
-\frac{8}{27} (2 - 2\,x + x^2)
 \Li_3(1-x)
-\frac{8}{27} (50 + 46\,x + 13\,x^2)
 \Li_3\left(-\frac{(1-x)}{x}\right)
\\&
-\frac{8}{9} (6 - 4\,x - x^2)
 \Li_2(1-x)
+\frac{16}{27} (19 + 17\,x + 5\,x^2)
 \Li_2(1-x)\ln(x)
\\&
+\frac{8}{81} (19 + 17\,x + 5\,x^2)
 \ln^3(x)
-\frac{4}{27} (18 - 10\,x - 9\,x^2)
 \ln^2(x)
\\&
-\frac{2}{27} (129 - 212\,x - 69\,x^2)
 \ln(x)
+\frac{16}{9} (4 + 4\,x + x^2)
 \zeta_2\ln(x)
\\&
+\frac{4}{27}(-86 + 53\,x + 33\,x^2)
+\frac{32}{9}(3 - 2\,x - x^2)
\,\zeta_2
\end{split}
\end{equation}
%
%
and
\begin{equation}
\begin{split}
\Delta_{qqA}^{F}&(x)= 0\,.
\end{split}
\end{equation}


For the scattering of a quark--(anti-)quark pair of distinct flavor, we
find
\begin{equation}
\begin{split}
\Delta_{qq'A}^{A}&(x)=
-\frac{32}{9} (3 - 2\,x - x^2)
 \ln^2(1-x)
-\frac{16}{9} (4 + 4\,x + x^2)
 \ln^2(1-x)\ln(x)
\\&
+\frac{4}{3} (17 - 12\,x - 5\,x^2)
 \ln(1-x)
+\frac{8}{9} (12 - 8\,x - 5\,x^2)
 \ln(1-x)\ln(x)
\\&
+\frac{8}{9} (4 + 4\,x + x^2)
 \ln(1-x)\ln^2(x)
-\frac{32}{9} (4 + 4\,x + x^2)
 \ln(1-x)\Li_2(1-x)
\\&
-\frac{32}{9} (4 + 4\,x + x^2)
 \Li_3\left(-\frac{(1-x)}{x}\right)
\\&
-\frac{8}{9} (6 - 4\,x - x^2)
 \Li_2(1-x)
+\frac{8}{3} (4 + 4\,x + x^2)
 \Li_2(1-x)\ln(x)
\\&
+\frac{4}{9} (4 + 4\,x + x^2)
 \ln^3(x)
-\frac{8}{9} (3 - 2\,x - x^2)
 \ln^2(x)
\\&
-\frac{2}{9} (43 - 68\,x - 29\,x^2)
 \ln(x)
+\frac{16}{9} (4 + 4\,x + x^2)
 \zeta_2\,\ln(x)
\\&
-\frac{2}{9}(61 - 46\,x - 15\,x^2)
+\frac{32}{9}(3 - 2\,x - x^2)
\,\zeta_2
\end{split}
\end{equation}
%
%
and
\begin{equation}
\begin{split}
\Delta_{qq'A}^{F}&(x)= 0\,.
\end{split}
\end{equation}


For quark--anti-quark scattering (same quark flavor), we have
\begin{equation}
\begin{split}
\Delta_{q\bar qA}^{A}&(x)=
-\frac{32}{81} (14 + 21\,x - 48\,x^2 + 13\,x^3)
 \ln^2(1-x)
-\frac{16}{9} (4 + 4\,x + x^2)
 \ln^2(1-x)\ln(x)
\\&
+\frac{4}{81} (75 + 892\,x - 1351\,x^2 + 384\,x^3)
 \ln(1-x)
\\&
+\frac{8}{81} (76 + 72\,x - 189\,x^2 + 64\,x^3)
 \ln(1-x)\ln(x)
\\&
+\frac{8}{9} (4 + 4\,x + x^2)
 \ln(1-x)\ln^2(x)
-\frac{32}{9} (4 + 4\,x + x^2)
 \ln(1-x)\Li_2(1-x)
\\&
-\frac{40}{27} (2 + 2\,x + x^2)
 \Li_3(1-x)
-\frac{8}{9} (18 + 18\,x + 5\,x^2)
 \Li_3\left(-\frac{(1-x)}{x}\right)
\\&
+\frac{10}{27} (2 + 2\,x + x^2)
 \Li_3(1-x^2)
+\frac{2}{9} (2 + 2\,x + x^2)
 \Li_3\left(-\frac{(1-x^2)}{x^2}\right)
\\&
+\frac{8}{27} (2 + 2\,x + x^2)
\left[\Li_3\left(\frac{1-x}{1+x}\right)
-\Li_3\left(-\frac{1-x}{1+x}\right)\right]
\\&
-\frac{8}{81} (12 + 30\,x - 93\,x^2 - 26\,x^3)
 \Li_2(1-x)
+\frac{16}{9} (7 + 7\,x + 2\,x^2)
 \Li_2(1-x)\ln(x)
\\&
-\frac{8}{27}\, (2\,x + x^2 + 6\,x^3)
 \Li_2(1-x^2)
-\frac{4}{9} (2 + 2\,x + x^2)
 \Li_2(1-x^2)\ln(x)
\\&
+\frac{8}{27} (5 + 5\,x + x^2)
 \ln^3(x)
-\frac{8}{81} (27 + 27\,x - 81\,x^2 + 44\,x^3)
 \ln^2(x)
\\&
+\frac{2}{81} (9 - 1064\,x + 2111\,x^2 - 768\,x^3)
 \ln(x)
+\frac{16}{9} (4 + 4\,x + x^2)
 \zeta_2\,\ln(x)
\\&
+\frac{20}{81}(101 - 462\,x + 520\,x^2 - 159\,x^3)
+\frac{16}{81}(11 + 93\,x - 147\,x^2 + 43\,x^3)
\,\zeta_2
\end{split}
\end{equation}
%
%
and
\begin{equation}
\begin{split}
\Delta_{q\bar qA}^{F}&(x)=
\frac{32(1-x)^3}{81}
\ln(1-x)
-\frac{16(3 - 9\,x + 12\,x^2 - 4\,x^3)}{81}
\ln(x)
\\&
-\frac{8(41 - 111\,x + 111\,x^2 - 41\,x^3)}{243}
\,.
\end{split}
\end{equation}


In the expressions above, we have identified the renormalization and the
factorization scale with the Higgs boson mass, $\mu_R=\mu_F=\mahiggs$.
The dependence on these scales is logarithmic and can readily be
reconstructed by employing scale invariance of the total partonic cross
section. 

\end{appendix}

\def\app#1#2#3{{\it Act.~Phys.~Pol.~}{\bf B #1} (#2) #3}
\def\apa#1#2#3{{\it Act.~Phys.~Austr.~}{\bf#1} (#2) #3}
\def\annphys#1#2#3{{\it Ann.~Phys.~}{\bf #1} (#2) #3}
\def\cmp#1#2#3{{\it Comm.~Math.~Phys.~}{\bf #1} (#2) #3}
\def\cpc#1#2#3{{\it Comp.~Phys.~Commun.~}{\bf #1} (#2) #3}
\def\epjc#1#2#3{{\it Eur.\ Phys.\ J.\ }{\bf C #1} (#2) #3}
\def\fortp#1#2#3{{\it Fortschr.~Phys.~}{\bf#1} (#2) #3}
\def\ijmpc#1#2#3{{\it Int.~J.~Mod.~Phys.~}{\bf C #1} (#2) #3}
\def\ijmpa#1#2#3{{\it Int.~J.~Mod.~Phys.~}{\bf A #1} (#2) #3}
\def\jcp#1#2#3{{\it J.~Comp.~Phys.~}{\bf #1} (#2) #3}
\def\jetp#1#2#3{{\it JETP~Lett.~}{\bf #1} (#2) #3}
\def\mpl#1#2#3{{\it Mod.~Phys.~Lett.~}{\bf A #1} (#2) #3}
\def\nima#1#2#3{{\it Nucl.~Inst.~Meth.~}{\bf A #1} (#2) #3}
\def\npb#1#2#3{{\it Nucl.~Phys.~}{\bf B #1} (#2) #3}
\def\nca#1#2#3{{\it Nuovo~Cim.~}{\bf #1A} (#2) #3}
\def\plb#1#2#3{{\it Phys.~Lett.~}{\bf B #1} (#2) #3}
\def\prc#1#2#3{{\it Phys.~Reports }{\bf #1} (#2) #3}
\def\prd#1#2#3{{\it Phys.~Rev.~}{\bf D #1} (#2) #3}
\def\pR#1#2#3{{\it Phys.~Rev.~}{\bf #1} (#2) #3}
\def\prl#1#2#3{{\it Phys.~Rev.~Lett.~}{\bf #1} (#2) #3}
\def\pr#1#2#3{{\it Phys.~Reports }{\bf #1} (#2) #3}
\def\ptp#1#2#3{{\it Prog.~Theor.~Phys.~}{\bf #1} (#2) #3}
\def\ppnp#1#2#3{{\it Prog.~Part.~Nucl.~Phys.~}{\bf #1} (#2) #3}
\def\sovnp#1#2#3{{\it Sov.~J.~Nucl.~Phys.~}{\bf #1} (#2) #3}
\def\sovus#1#2#3{{\it Sov.~Phys.~Usp.~}{\bf #1} (#2) #3}
\def\tmf#1#2#3{{\it Teor.~Mat.~Fiz.~}{\bf #1} (#2) #3}
\def\tmp#1#2#3{{\it Theor.~Math.~Phys.~}{\bf #1} (#2) #3}
\def\yadfiz#1#2#3{{\it Yad.~Fiz.~}{\bf #1} (#2) #3}
\def\zpc#1#2#3{{\it Z.~Phys.~}{\bf C #1} (#2) #3}
\def\ibid#1#2#3{{ibid.~}{\bf #1} (#2) #3}

\newcommand{\jref}[3]{{\bf #1}, #3 (#2)}

\end{document}